# Datensouveränität für Verbraucher:innen: Technische Ansätze durch KI-basierte Transparenz und Auskunft im Kontext der DSGVO

Elias Grünewald, Frank Pallas

Information Systems Engineering, Technische Universität Berlin
*{gruenewald, frank.pallas}@tu-berlin.de*

**Abstract.** Hinreichende Datensouveränität gestaltet sich für Verbraucher:innen in der Praxis als äußerst schwierig. Die Europäische Datenschutzgrundverordnung garantiert umfassende Betroffenenrechte, die von verwantwortlichen Stellen durch technisch-organisatorische Maßnahmen umzusetzen sind. Traditionelle Vorgehensweisen wie die Bereitstellung länglicher Datenschutzerklärungen oder der ohne weitere Hilfestellungen angebotene Download von personenbezogenen Rohdaten werden dem Anspruch der informationellen Selbstbestimmung nicht gerecht. Die im Folgenden aufgezeigten neuen technischen Ansätze insbesondere KI-basierter Transparenz- und Auskunftsmodalitäten zeigen die Praktikabilität wirksamer und vielseitiger Mechanismen. Hierzu werden die relevanten Transparenzangaben teilautomatisiert extrahiert, maschinenlesbar repräsentiert und anschließend über diverse Kanäle wie virtuelle Assistenten oder die Anreicherung von Suchergebnissen ausgespielt. Ergänzt werden außerdem automatisierte und leicht zugängliche Methoden für Auskunftsersuchen und deren Aufbereitung nach Art. 15 DSGVO. Abschließend werden konkrete Regulierungsimplikationen diskutiert.

# 1 Einleitung

Die Europäische Datenschutzgrundverordnung (DSGVO) (Europäische Union, 2016) beeinflusst die datenschutzfreundliche Gestaltung sozio-technischer Systeme weltweit (Rustad und Koenig, 2019). Insbesondere werden zentrale Verbraucher:innen bzw. Betroffenenrechte (und Pflichten der für die Verarbeitung personenbezogener Daten Verantwortlichen und deren Auftragsverarbeiter:innen) hinsichtlich Transparenz und Auskunft reguliert. Unter *Transparenz* werden nachfolgend die umfassenden Informationspflichten verstanden, die Verbraucher:innen in präziser und leicht verständlicher Form zeitnah zu übermitteln sind (Art. 12–14 DSGVO). Diese beinhalten insbesondere Angaben über die mögliche Art, Zweck, Dauer oder Rechtsgrundlage der Verarbeitung personenbezogener Daten im Kontext eines Dienstangebots. Traditionell erfolgen diese Angaben über eine Datenschutzerklärung. Als *Auskunft* wird demgegenüber die Bereitstellung der speziell zu einer betroffenen Person verarbeiteten Daten bezeichnet. Art. 15 DSGVO definiert hier eine Pflicht der verantwortlichen Stelle zur Herausgabe ebendieser Daten, sofern sie die Freiheiten und Rechte anderer Personen nicht einschränken. Art. 20 DSGVO statuiert – primär mit dem Ziel der Übertragbarkeit – verwandte Pflichten zur Herausgabe von Daten und schreibt hierfür die Verwendung strukturierter, maschinenlesbarer Formate vor. In der Praxis werden mehrheitlich sogenannte persönliche Dashboards oder – nicht zuletzt auch zur Erfüllung weiterer Pflichten z. B. aus Art. 20 DSGVO – der Download von großen Dateiarchiven angeboten.

Beide genannten Schwerpunkte von Betroffenenrechten erfüllen in ihrer derzeitigen technischen Umsetzung den eigentlichen Zweck der Datensouveränität weitesgehend nicht. Datenschutzerklärungen werden von Verbraucher:innen nachweislich mehrheitlich weder gelesen noch verstanden (Fabian et al., 2017). Ferner sind viele Dashboards unvollständig (Bier et al., 2016) und das Erfassen des Inhalts und der Tragweite der in den Dateiarchiven enthaltenen Informationen ist nur mit hoher Data Literacy möglich (Alizadeh et al., 2020).

Nachfolgend werden deshalb neuartige technische Ansätze zur Datensouveränität vorgestellt, die zeigen wie

1. Transparenzinformationen zuerst maschinenlesbar repräsentiert und danach durch KI-basierte Technologien wie interaktive Chatbots, virtuelle Sprachassistenten oder kontextsensitive Browser-Erweiterungen ausgespielt und
2. Auskunftsersuchen durch Verbraucher:innen technikgestützt und damit möglichst aufwandsarm eingeholt und beispielhaft automatisiert aufbereitet werden können.

Abschließend werden mögliche regulatorische Implikationen zur breiten Etablierung neuer, technisch vermittelter Ansätze und Verfahren im Kontext von Transparenz- und Auskunftsrechten diskutiert.

# 2 Transparenz

Transparenz ist seit Jahrzehnten ein Kernprinzip des Datenschutzes. Im Europäischen Datenschutzrecht ist die transparente Verarbeitung personenbezogener Daten mehrfach verankert. Art. 5 Abs. 1 lit. a DSGVO betont die Nachvollziehbarkeit und kollokiert Transparenz mit den Prinzipien der Rechtmäßigkeit und einer Verarbeitung nach Treu und Glauben (Fairness). Des Weiteren verpflichtet Art. 12 DSGVO (u. a. i. V. m. Erwägungsgrund 58) die für die Verarbeitung verantwortlichen Stellen zur transparenten Information, unverzüglichen Kommunikation und zugänglichen Modalitäten. In Verbindung mit den Art. 13 und 14 DSGVO gehen dann umfassende Informationspflichten (sogenannte Transparenzinformationen) einher. Neben allgemeinen, auf den Dienst bezogenen, Angaben (wie u. a. Kontaktdaten des oder der Verantwortlichen und Datenschutzbeauftragten) werden auch detaillierte Informationen über die jeweilige Rechtsgrundlage, Zweckbestimmung, Empfänger:innen bzw. involvierte Drittparteien für die Verarbeitung jeder personenbezogenen Datenkategorie verlangt. Diese Betroffenenrechte werden üblicherweise durch die Bereitstellung textueller Datenschutzerklärungen erfüllt. Diverse verwandte Arbeiten belegen jedoch, dass diese häufig durch die Verwendung juristischer Fachsprache, ihre Länge und Zugänglichkeit oder die inhärente Komplexität nicht oder nur teilweise gelesen und verstanden werden (McDonald und Cranor, 2008; Pollach, 2007; Torre et al., 2020). Insbesondere fällt es betroffenen Personen schwer, die für sie individuell eintretenden Folgen bei der Benutzung eines Dienstes zu verstehen (Chang et al., 2018). Problematisch ist weiter die bewusste Verschleierung oder beförderte Verwirrung über die Folgen einer Einwilligung in die Verarbeitung personenbezogener Daten. In der UI/UX-Literatur wird diese Praxis mangelnder Gebrauchstauglichkeit als „Dark pattern" bezeichnet (Nouwens et al., 2020).

Infolgedessen ist der Status quo datenschutzrechtlicher Transparenz unzureichend. Hinzu kommt, dass die Architekturen verteilter Systeme – also die technische Grundlage aller internetbasierten Dienste mit Millionen Kund:innen – zunehmend komplexer werden. Um moderne, also dynamische und skalierbare Dienste anbieten zu können, werden immer häufiger die Methoden des nativen Cloud Computings eingesetzt (Kratzke und Quint, 2017). Hierzu gehört die flexible Zusammenstellung von Microservices (gekapselte Funktionalitäten als komponentenorientiere Teilsysteme), die über automatisierbare Programmierschnittstellen (APIs) kommunizieren (Cloud Native Computing Foundation (CNCF), 2018). Der überwiegende Teil der Verarbeitung personenbezogener Daten findet dann innerhalb der Cloud-Infrastruktur (z. B. virtuelle Maschinen, vorkonfigurierte Datenbanken, bereits trainierte Machine-Learning-Angebote usw.) eines oder mehrerer Anbieter (wie Amazon Web Services oder Google Cloud Platform) statt. Hierbei steigt auch die Komplexität der datenschutzrechtlichen Beurteilung: Durch die Nutzung der bereitgestellten Rechenleistung, Speicher und Netzwerkkomponenten ergibt sich beispielsweise die Begründung einer Auftragsdatenverarbeitung oder ausreichend abzusichernder Drittlandtransfers, sofern die gemieteten Angebote außerhalb der Europäischen Union (oder in Staaten mit angemessenem Schutzniveau, vgl. Art. 45ff. DSGVO) betrieben werden. Ferner werden solche IT-Systeme agil entwickelt. Mithilfe schneller Entwicklungszyklen und durch einen hohen Grad von Automatisierung bei der Bereitstellung von Software (Continuous Integration / Continuous Delivery) ändern sich die Bedingungen der personenbezogenen Datenverarbeitung kontinuierlich (Grünewald et al., 2021). Hierbei stoßen die etablierten Modalitäten zur Herstellung von Transparenz erneut an ihre Grenzen: Verantwortlichen Stellen fehlt es an aufwandsarm zu integrierenden Tools zur Überwachung der Verarbeitung und betroffene Personen sind weiterhin schlecht informiert. Zudem gestaltet sich die unabhängige externe Kontrolle, Begutachtung oder Zertifizierung – etwa durch Aufsichtsbehörden oder Zertifizierungsstellen – wegen hochgradig spezifischer Software-Architekturen als äußerst schwierig. Deshalb werden im Folgenden einige neue Ansätze zur Verbesserung datenschutzrechtlicher Transparenz diskutiert. Zuerst wird hierfür von bereits flächendeckend vorliegenden Datenschutzerklärungen ausgegangen.

## 2.1 Teilautomatisierte Extraktion relevanter Informationen aus Datenschutzerklärungen

Vorbedingung für diverse Anwendungen im Kontext Transparenz ist ein hinreichend großer Datenbestand an Transparenzinformationen von realen, für die Verarbeitung von personenbezogenen Daten verantwortlichen Stellen. Hierzu ist die systematische Erfassung der in Datenschutzerklärungen kodifizierten Informationen naheliegend. Während verwandte Arbeiten Datenschutzerklärungen als Ganzes katalogisieren und Veränderungen über die Zeit nachvollziehen (Wilson et al., 2016), geht der im Folgenden beschriebene – im Rahmen des BMJV-geförderten Projekts DaSKITA[1] umgesetzte – Ansatz weiter: Alle gemäß Art. 13 und 14 DSGVO zwingend anzugebenen Informationen werden hierzu auf Detailebene untersucht.

Die manuelle Erstellung eines solchen Korpus durch einzelne (notwendigerweise fachkundige) Personen ist weder praktikabel noch zugänglich für eine ggf. nachfolgende Analyse oder Weiternutzung der erkannten Angaben. Deshalb wird der Korpus von Transparenzinformationen mithilfe einer eigens im Kontext entwickelten Annotationsplattform erstellt werden. Diese wird als Open-Source-Software zur freien weiteren Verwendung mitsamt Dokumentation bereitgestellt.[2] Hierdurch können bereits jetzt teilautomatisiert relevante Informationen markiert und in einer offenen Datenbank abgespeichert werden.

---

[1] https://www.ise.tu-berlin.de/daskita und https://daskita.github.io
[2] https://github.com/DaSKITA/tilter

Innerhalb der Plattform werden den Annotator:innen *(i)* Datenschutzerklärungen präsentiert. Für jede Erklärung werden die Annotator:innen dann *(ii)* Schritt für Schritt nach dem Vorliegen aller relevanten Angaben befragt und *(iii)* zur Markierung der entsprechenden Textstelle(n) aufgefordert. Durch eine aspektorientierte Aufteilung ist auch eine kollaborative und zeitgleiche Annotation durch mehrere Expert:innnen möglich. Resultat der Annotation ist eine Übersicht aller angegebenen bzw. fehlender Informationen im Kontext datenschutzrechtlicher Transparenz. Somit ist auch eine vorläufige Bewertung der jeweils betrachteten Datenschutzerklärung zumindest im Ansatz möglich, wenngleich diese Einschätzung offenkundig nicht vollends die Richtig- und Vollständigkeit prüfen kann.

In einem zweiten Schritt soll die Extraktion der relevanten Informationen auch vollautomatisiert erfolgen können. Hierzu werden u. a. KI-basierte Methoden der *Named Entity Recognition* und des *Natural Language Processing* eingesetzt werden (Dozier et al., 2010). Entsprechende Arbeiten befinden sich zurzeit in Vorbereitung. Durch die oben beschriebene manuelle Annotation wurde bereits ein erster Korpus erstellt, der laufend erweitert und – wie die Annotationsplattform – bereits öffentlich zugänglich ist.[3] Die Testung und das Training der genannten lernenden Verfahren kann durch das Vorliegen der Annotationen optimiert werden; andererseits kann auch die manuelle Annotation durch erste Ergebnisse der KI-basierten Extraktion erleichtert werden (z. B. durch Vorschläge zur Annotation). Diese Prozesse werden in Abb. 1 schematisch dargestellt.

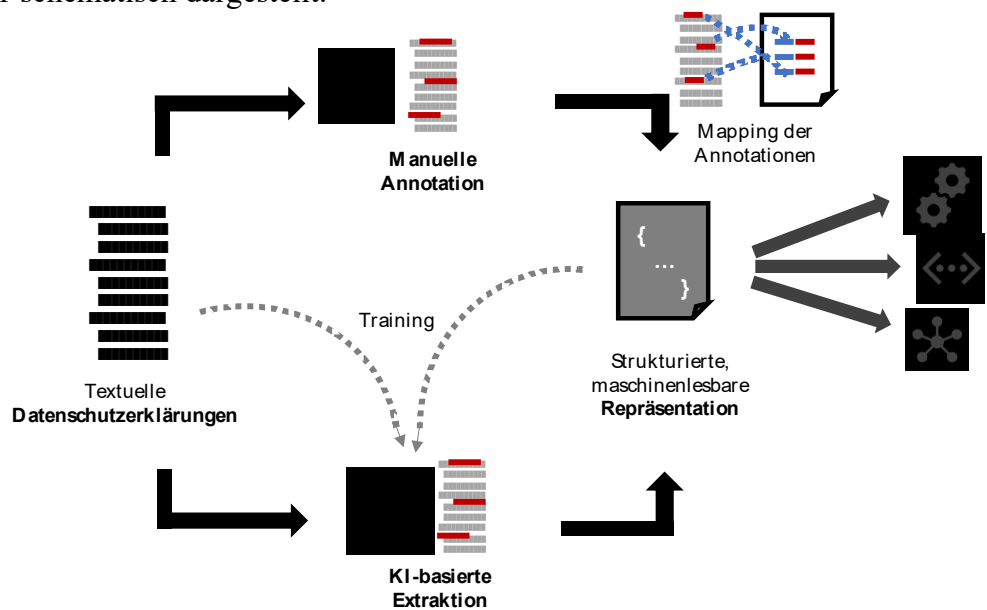

Abbildung 1. Vorgehen zur Erstellung eines Korpus maschinenlesbarer Transparenzinformationen.

---

[3] https://github.com/Transparency-Information-Language/tilt-corpus

Die extrahierten Ergebnisse werden anschließend automatisch in eine strukturierte, maschinenlesbare Repräsentation überführt. Diese wird im Folgenden näher beschrieben.

## 2.2 Strukturierte Repräsentation und Bereitstellung maschinenlesbarer Transparenzinformationen

Technische Ansätze zur datenschutzfreundlichen Transparenz werden als *Transparency Enhancing Technologies* (TETs) klassifiziert (Zimmermann, 2015). Die strukturierte Repräsentation von Transparenzinformationen, basierend auf einer formalen Sprachspezifikation, ermöglicht eine Vielzahl möglicher TETs. TILT (Transparency Information Language and Toolkit) bildet die Grundlage für jene Entwicklungen (Grünewald und Pallas, 2021). Das Framework wurde entlang der technischen und rechtlichen Anforderungen für praxistaugliche privatheitswahrende Technikgestaltung (Privacy Engineering) entwickelt. Für die DSGVO-konforme Ausrichtung auf Transparenz wurde hierzu eine detaillierte Modellierung der notwendigen Ausdrucksmächtigkeit entlang der Art. 12, 13, 14, 15 und 30 vorgenommen. Durch den Abgleich mit mehreren Dutzend real existierenden Datenschutzerklärungen konnte außerdem nachgewiesen werden, dass alle maßgeblichen Informationen mithilfe der neuartigen Repräsentation abgebildet werden können. Durch die Nutzung gängiger Open Source-Technologien wie JSON und JSON Schema (Pezoa et al., 2016) zur Implementierung der Sprache konnte außerdem ein hoher Grad an Erweiterbarkeit erzielt werden. Die leichte Integration in existierende Systemarchitekturen wird durch begleitende Programmierbibliotheken ebenfalls sichergestellt.[4]

Die oben vorgestellte Annotationsplattform ist ebenfalls Teil des an TILT orientierten technologischen Ökosystems. So werden die gespeicherten Annotationen automatisch in das maschinenlesbare Format überführt. Die so gewonnenen Transparenzinformationen können zugleich über verschiedene Programmierschnittstellen (REST und GraphQL) aus einer offenen Datenbank abgerufen werden.[5] Alle vorgenannten technischen Werkzeuge richten sich explizit an die verantwortlichen Stellen und deren Auftragsdatenverarbeiter:innen. Ergänzt werden die bereits genannten Möglichkeiten durch Projekte, die sich auch an Nicht-Techniker:innen richten und die Erstellung maschinenlesbarer Transparenzinformationen auch für Laien zugänglich machen können. Exemplarisch sei ein prototypischer, Baukasten-basierter Editor zur Erstellung von TILT-Dokumenten[6] erwähnt. Diese Werkzeuge sollen das Informationsmanagement und die Kommunikation aller Beteiligten innerhalb der verantwortlichen Stelle erleichtern und bereichern.

---

[4] https://github.com/Transparency-Information-Language/meta

[5] https://github.com/Transparency-Information-Language/tilt-hub

[6] https://github.com/ProgPrak2021/PeachPrivacy

Diese neuartigen Ansätze ermöglichen die darauf aufbauende Entwicklung zahlreicher Ausspielungsmöglichkeiten, die dann Verbraucher:innen maßgeblich bei der Wahrnehmung ihrer Rechte zur informationellen Selbstbestimmung unterstützen. Ausgewählte Beispiele solcher benutzer:innenorienten Technologien werden im Folgenden beschrieben.

## 2.3 Kontextsensitive und präferenzorientiere Ausspielung von Transparenzinformationen

Im Unterschied zu klassischen, rein textuellen Datenschutzerklärungen rücken im Rahmen dieser Arbeit kontextsensitive und präferenzorientierte Modalitäten in den Vordergrund. Durch die zunehmende Nutzung mobiler Endgeräte oder auch durch die Etablierung eines Internet of Things (z. B. im Smart Home) müssen neue Präsentationsformen entwickelt werden. Während bei einer klassischen Desktop-Oberfläche eine textuelle Erklärung (noch) rezipierbar wäre, besteht diese Möglichkeit beispielsweise bei der Nutzung eines üblichen Smart Speakers im Verbraucher:innenhaushalt nicht, da oft keine grafische Oberfläche bzw. kein Display vorhanden ist (vgl. für den Aspekt der Einwilligung auch Ulbricht und Pallas 2018). Darüber hinaus besteht in der Bevölkerung ein unterschiedlich hoher Grad an Digital Literacy (Park, 2013). Demnach müssen auch kompetenz- und präferenzorientierte Methoden entwickelt werden, die den unterschiedlichen Informationsbedarfen durch passende Detail- oder Überblicksangaben gerecht werden können (Schaub et al., 2015). Drei exemplarische Ausspielungsmöglichkeiten, die auf den mittels TILT repräsentierten und zugänglich gemachten Daten basieren, werden im Folgenden kurz vorgestellt.

### 2.3.1 Chatbot und Virtual assistant

Durch KI-basierte Aufbereitung können datenschutzrechtliche Transparenzinformationen für Verbraucher:innen mit einem interaktiven digitalen Assistenten[7] bereitgestellt werden. Die Forschung ist damit direkt anschlussfähig an vergleichbare Ergebnisse zur intelligenten Dialogführung (Conversational AI) mittels Chatbots (textbasiert) oder virtueller Assistenten (sprachbasiert) (Harkous et al., 2016, 2018). Der erste hiermit veröffentlichte Prototyp für den virtuellen Assistenten Amazon Alexa und den web-basierten Chat kann somit bereits Fragen zur Verarbeitung personenbezogener Daten beantworten, die sonst nur durch aufwändige Recherche innerhalb einer textuellen Datenschutzerklärung zu bewältigen wären. Dabei können direkt Informationen zu mehreren Diensten bereitgestellt werden, ohne den Kanal zu verlassen. Das System basiert auf dem quelloffenen Natural-Language-Understanding-Framework Rasa (Bocklisch et al., 2017). Hierbei wird auch ein Modell trainiert, um mit lernenden Verfahren die passenden Dialogflüsse zu verbessern. Für die Ausgabe der Informationen wird ein dedizierter Webserver mit offener API bereitgestellt. Darüber hinaus erfolgt derzeit eine Nutzer:innenstudie zur Evaluierung des Ansatzes. Die zukünftige Entwicklung von TIBO (Transparency Information BOt) soll auch Funktionalitäten zur teilautomatisierten Auskunft berücksichtigen.

---

[7] https://github.com/DaSKITA/chatbot

2.3.2 Browser-Erweiterung als Privacy-Agent

Eine weitere Ausspielungsmöglichkeit wurde für den am weitesten verbreiteten Internet-Browser Google Chrome implementiert. Die Idee von Privatsphäre-Agenten wurde bereits breit in der Literatur diskutiert (Cranor et al., 2006). Frühe Arbeiten auf Basis von P3P (Cranor, 2002), wie beispielsweise das Plugin PrivacyBird (Vu et al., 2010), erlangten hohe Aufmerksamkeit. Nachdem nun die Gesetzgebung explizit auch Bildsymbole (Privacy Icons; vgl. Rossi und Palmirani 2020; Habib et al. 2021) vorsieht (vgl. Art. 12 Abs. 7 DSGVO) und mit TILT eine maschinenlesbare Repräsentation mit ausreichender Ausdrucksmächtigkeit bereitsteht, wird eine browserbasierte Kommunikation von Transparenzinformationen nun praktikabel. Prototypisch wurde hierfür bereits ein Plugin implementiert, das eine Kurzzusammenfassung relevanter Angaben zur jeweils besuchten Seite darstellen kann.[8] Diese enthält u. a. Angaben zum oder zur Verantwortlichen, der Anzahl Drittlandtransfers oder ob automatisierte Entscheidungssysteme im Einsatz sind. Eine zukünftige Integration mit Transparenz- und Einwilligungsframeworks (Matte et al., 2020), die v. a. im Zusammenhang mit sogenannten Cookie-Bannern verwendet werden, liegt nahe.

2.3.3 Integration in Suchergebnisse

Transparenz ermöglicht, informierte Entscheidungen von Verbraucher:innen u. a. hinsichtlich der Wahl einen Dienst zu nutzen oder nicht. Das Notice-and-Choice-Modell (Cate, 2010) sieht demnach vor, dass betroffene Personen vor (ex ante) der tatsächlichen Verarbeitung personenbezogener Daten die entsprechende Folgenabschätzung treffen können. In der Praxis kommen allerdings oftmals bereits beim ersten Besuch einer Website Tracking-Methoden wie Third-Party-Cookies (Cahn et al., 2016) oder Browser-Fingerprinting (Pugliese et al., 2020) zum Einsatz. Ein wirksamer Schutz ist dann nur eingeschränkt möglich und insbesondere eher Menschen mit einem fortgeschrittenen Wissen über etwaige Gefahren vorbehalten. Deshalb wird im Folgenden eine eigens entwickelte Browser-Erweiterung vorgestellt, die es Verbraucher:innen ermöglicht, einen schnellen Überblick über bekannte Tracking-Mechanismen und potentiell unerwünschte Verarbeitungen personenbezogener Daten zu erhalten, bevor eine Website tatsächlich geöffnet wird.

---

[8] https://github.com/Transparency-Information-Language/chrome-tilt

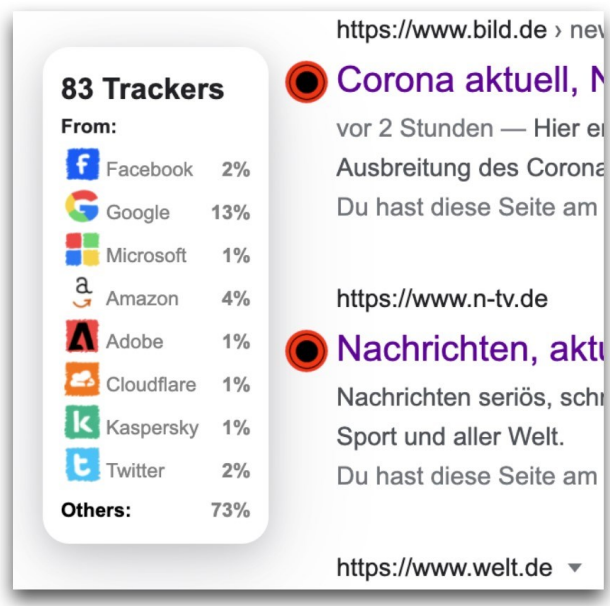

Abbildung 2. Exemplarische Privacy Labels.

Im Kern werden dazu die Suchergebnisse der verbreitetsten Suchmaschine Google mit Transparenzinformationen angereichert. Bereits 2011 (Tsai et al., 2011) wurde in einer verwandten Studie nachgewiesen, dass Nutzer:innen so einerseits besser informiert sind und darüber hinaus sogar bereit sind, höhere Preise für Produkte zu bezahlen, wenn die Dienste einen höheren Grad gewahrter Privatheit signalisieren. Dies wird vor allem mit einem verstärkt beigemessenem Vertrauen erklärt. Die entwickelte Erweiterung platziert Privacy Labels, die – je nach Einstellung – entweder als Ampelsymbol (vgl. Abb. 2), Kurzzusammenfassung oder durch einen Scoring-Verfahren die entsprechenden Informationen bereitstellen. Grundlage der Risikoberechnung ist hierbei die Integration diverser Datenbanken und APIs, die detaillierte Informationen über tausende der beliebtesten Websites bereitstellen: *WhoTracksMe* (Karaj et al., 2018), *Phishstats*[9], *PrivacySpy*[10], *Google Safe Browsing API*[11], *Terms of Service Didn't Read*[12] und TILT-Dokumente via *tilt-hub*[13]. Die Browser-Erweiterung berechnet zudem lokale Statistiken, um das eigene Browsing-Verhalten kontinuierlich überprüfen zu können. Durch diverse Einstellungsmöglichkeiten können Verbraucher:innen die Transparenzinformationen nach eigenen Präferenzen justieren. Die gesamte Erweiterung ist ebenfalls als Open-Source-Software verfügbar[14] und speichert keine individuellen Suchprofile. Mögliche zukünftige Entwicklungen beinhalten insbesondere die Portierung auf mobile Browser, die Integration weiterer Datenquellen oder auch die Möglichkeit zur Neusortierung von Suchergebnissen zur schnelleren Auffindbarkeit datenschutzfreundlicher Dienste.

---

[9] https://phishstats.info/

[10] https://github.com/politiwatch/privacyspy

[11] https://developers.google.com/safe-browsing

[12] https://tosdr.org/

[13] https://github.com/Transparency-Information-Language/tilt-hub

[14] https://github.com/ProgPrak2021/CODE

# 3 Auskunft

Die vorangegangen Beispiele neuer Möglichkeiten zur technikgestützten Erfüllung von Transparenzpflichten informieren Verbraucher:innen über generelle Angaben zur Erhebung und Verarbeitung personenbezogener Daten. Art. 15 sowie Art. 20 DSGVO erweitern diese Betroffenenrechte um den Anspruch auf Kopie bzw. Export und ggf. Übertragbarkeit der konkret auf das Individuum bezogenen Daten. Relevante regulatorische Rahmenwerke wie der California Consumer Privacy Act (CCPA) definieren ebenfalls ein vergleichbares „Right to Access" (Veys et al., 2021). In der Praxis ist das Recht auf Auskunft allerdings bislang mangelhaft implementiert. Als Hürden können beispielhaft aufgezählt werden (Alizadeh et al., 2020):

- Verbraucher:innen sind unzureichend über die Existenz des Rechts auf Auskunft informiert.
- Die Auffindbarkeit der Methoden zur Einholung der Auskunft ist unzureichend.
- Die Übermittlung der Auskunft erfolgt oftmals derart zeitverzögert, dass das Interesse daran ggf. verloren geht oder die Informationen nicht mehr aktuell sind.
- Die übermittelten Daten sind in vielen Fällen nicht rezipierbar, ohne ein entsprechendes Fachwissen zur Aufbereitung großer Rohdatenbestände zu haben.
- Das Recht auf Datenübertragbarkeit ist in der praktischen Umsetzung (mit wenigen Ausnahmen[15]) mangels ökonomischer Anreize (Syrmoudis et al., 2021), wegen der fehlenden Implementierung von Import-Funktionalitäten (Syrmoudis et al., 2021) und aufgrund technischer Schwierigkeiten bei der Abbildung der Datenbestände verschiedener Verantwortlicher nicht umgesetzt (Pandit et al., 2020).

Im Folgenden werden deshalb zwei technische Ansätze vorgestellt, die einerseits die Einholung und danach die Aufbereitung personenbezogener Daten für Verbraucher:innen erleichtern können.

---

[15] Bekannte Dienstanbieter arbeiten zwar in einem Konsortium an einem *Data Transfer Project* (https://datatransferproject.dev/), für Verbraucher:innen sind diese Funktionalitäten allerdings weiterhin nicht verfügbar.

## 3.1 Einholung

Die Einholung einer Auskunft (Data Subject Access Request) erfolgt gewöhnlicherweise über eine manuelle Email-Anfrage, ein dediziertes Privacy Dashboard oder (in seltenen Fällen) über eine bereitgestellte API (Urban et al., 2019). Die Angaben zur Modalität sind gemäß Art. 12 DSGVO in transparenter Weise zur Verfügung zu stellen. Demnach ist die Auskunft in der Regel unentgeltlich und unverzüglich zu erteilen. Da die Modalitäten jedoch bei jedem Dienst spezifisch und somit mit hohen Transaktionsaufwänden (insbesondere Such- und Informationskosten) verbunden sind, nehmen bislang nur wenige Verbraucher:innen ihre diesbezüglichen Rechte wahr.

Um dem zu begegnen, wurde eine Browser-Erweiterung entwickelt, die die Einholung einer Auskunft für eine beliebige Anzahl von genutzten Diensten vereinheitlicht und damit enorm vereinfacht. DARA[16] (Data Subject Access Request & Analysis) wird hiermit als Open-Source-Erweiterung für Google Chrome und Mozilla Firefox zur Verfügung gestellt und bietet eine gebrauchstaugliche grafische Oberfläche zum Anfordern personenbezogener Daten (vgl. Abb. 3).

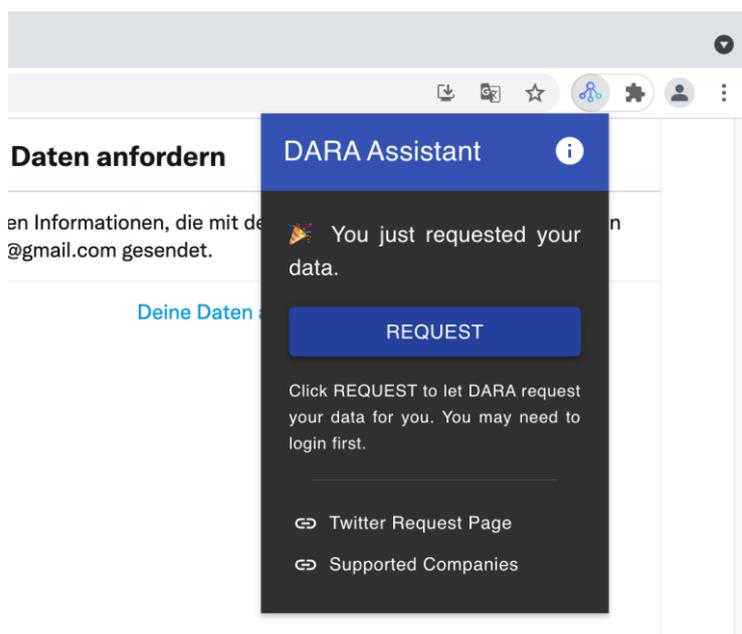

Abbildung 3. Auskunftsanfrage mittels DARA bei Twitter.

DARA integriert hierzu einen Datenbestand von JustGetMyData[17], der aufbereitete Beschreibungen und Zugangslinks zu Auskunftsmodalitäten dutzender Dienste bereitstellt, und erweitert diesen um auch per Crowd-Sourcing gewonnene, vertiefende Informationen zu notwendigen Klickpfaden etc. Diese Angaben werden in maschinenlesbarer Form über eine API bereitgestellt. Beim Navigieren auf einen Dienst, der von der betroffenen Person genutzt wird, erkennt die Browser-Erweiterung automatisch, ob eine entsprechende Beschreibung vorliegt. Je nach Kontext und vorliegenden Informationen können dann direkte Download-Links oder Klickpfade zur Verfügung gestellt werden.

---

[16] https://github.com/ProgPrak2021/DARA

[17] https://github.com/justgetmydata/jgmd

Darüber hinaus beinhaltet DARA eine erweiterbare *Engine* zur Automatisierung des gesamten Prozesses. Wo die hierfür notwendigen Informationen verfügbar sind, navigiert die Erweiterung – ausgehend von einem gültigen Login beim entsprechenden Dienst – automatisch zur entsprechenden Auskunftsschnittstelle, wartet ggf. die Bereitstellung der Daten ab und stellt diese schließlich den Verbraucher:innen ohne weitere manuelle Interaktion direkt zur Verfügung. Die Erweiterung kann durch die Community permanent ergänzt werden. Ein entsprechender *Call for participation* ruft in diesem Kontext versierte Entwickler:innen auf, weitere Konnektoren zu relevanten Diensten bereitzustellen. Hierfür ist keine tiefere Kenntnis über die interne Logik der Erweiterung notwendig, sondern nur die maschinenlesbare Dokumentation der für ein Auskunftsersuchen notwendigen Schritte. Anschließend ist die neue Funktionalität für alle Nutzer:innen der Erweiterung zugänglich. Die Privatsphäre der betroffenen Person wird durch die Erweiterung nicht beeinträchtigt. Sämtliche Berechtigungen werden explizit und ausschließlich lokal erteilt und verarbeitet, die abgerufenen personenbezogenen Daten verlassen das lokale Endgerät nicht und eine Nutzungsstatistik wird nicht erhoben. Darüber hinaus verfolgt DARA einen edukativen Ansatz und stellt Begleitmaterial zur Interpretation der erhaltenen Daten bereit. Somit leistet die Erweiterung einen nennenswerten Beitrag zur Erhöhung der Datensouveränität und Data Literacy von Verbraucher:innen.

## 3.2 Aufbereitung

Das Ergebnis einer Auskunftsanfrage ist meist ein Datenarchiv, das eine Vielzahl von Dokumenten beinhaltet. Diese Daten sind ohne einschlägige Kenntnisse nicht zu interpretieren. Stattdessen wäre ein aufwändiges Datenanalyseverfahren notwendig, um nachvollziehen zu können, welche inhaltlichen Erkenntnisse ein Dienstanbieter aus den vorliegenden Daten schließen kann. Gleichzeitig sind die Daten oftmals hoch sensitiv, sodass eine Weitergabe an eine Drittpartei zur Delegation der Aufbereitung in den meisten Fällen ausgeschlossen sein wird. Deshalb wird im Folgenden ein dezentral entworfenes Tool zur Aufbereitung vorgestellt, das Verbraucher:innen tiefere Einblicke in die vorliegenden Datenbestände erteilen kann.

MirrorMe[18] wurde entwickelt, um einen Überblick über alle personenbezogenen Daten zu geben, die bereits mit verschiedenen Unternehmen geteilt wurden. Hierzu wird ein lokales Profil auf Grundlage der beauskunfteten Daten erstellt. MirrorMe ist eine Desktop-Anwendung, die u. a. Visualisierungen wie aggregierte Datenbestände als Diagramme und Karten bereitstellt. Aktuell unterstützt das Tool die populären sozialen Netzwerke Reddit, Facebook und Instagram. Schließlich steht eine optionale Scoreboard-Funktion bereit, die einen numerischen Risikofaktor ermittelt, der den Gesamtumfang der Auskunft zusammenfasst. Somit können – nach Einwilligung – Verbraucher:innen mit anderen Interessierten vergleichen, wie umfangreich die persönliche Nutzung eines Dienstes ist – ohne dabei offenlegen zu müssen, welche Daten konkret vorhanden sind.

---

[18] https://github.com/ProgPrak2021/MirrorMe

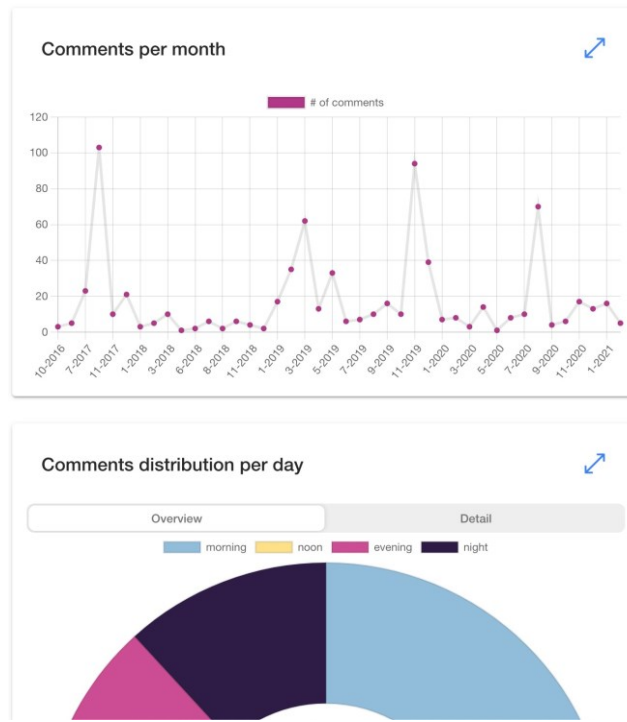

Abbildung 4. Auskunftsanalyse eines Reddit-Profils.

Insgesamt stellt MirrorMe einen vielversprechenden ersten Ansatz zur lokalen Aufbereitung von Auskunftsdaten dar. Denkbar sind etwa Erweiterungen, die tiefergehende Analysen der personenbezogenen Daten vornehmen. Solche Ergebnisse sind besonders deshalb relevant, weil davon ausgegangen werden muss, dass insbesondere die populärsten Dienste (wie soziale Netzwerke, aber auch Versicherungen, Fitness-Apps usw.) ein umfassendes Persönlichkeitsprofil erstellen oder zumindest theoretisch dazu in der Lage sind (Matz et al., 2020). Können Verbraucher:innen diese Praxis am eigenen Beispiel selbst nachvollziehen, verbessert dies fraglos ihr Verständnis der bestehenden Möglichkeiten und Risiken. Dies wiederum wirkt sich positiv auf ihre Fähigkeit zur tatsächlichen Ausübung von Datensouveränität und informationeller Selbstbestimmung aus.

# 4 Schlussbetrachtung

Die hier aufgezeigten Wege der technischen Umsetzung von Transparenz- und Auskunftsrechten sind vor allem als Ausgangspunkt besser informierter Diskussionen zu verstehen. Während besonders Dienstanbieter:innen dadurch auffallen, technologische Probleme und angeblich unüberwindbare Hürden zum effektiven Schutz von personenbezogenen Daten anzuführen, wird hier gezeigt, dass mit vergleichsweise geringem Implementierungsaufwand das Niveau der Datensouveränität signifikant gesteigert werden kann. Die aufgezeigten Wege erheben dennoch ausdrücklich keinen Anspruch auf Allgemeingültigkeit oder universelle Anwendbarkeit. Die komplexe Risikoabschätzung (Wuyts et al., 2020) im Kontext der Verarbeitung personenbezogener Daten muss weiterhin immer für den konkreten Anwendungsfall geschehen. Außerdem sollten stets auch *alle* Datenschutzprinzipien neben Transparenz und Auskunft bestmöglich adressiert werden.

Unabhängig davon zeigen die vorgestellten Ansätze einen Ausschnitt des sich eröffnenden technischen Möglichkeitsraums, der sich aus offenen Schnittstellen und Formaten, vereinheitlichten Standards sowie Open-Source-Entwicklungen von Privacy Enhancing Technologies (PETs) ergibt (Heurix et al., 2015). Insbesondere werden so auch Dienste mit verteilten Cloud-basierten Systemarchitekturen beleuchtet, bei denen die Komplexität der datenschutzrechtlichen und technischen Anforderungen stetig zunimmt.

Datenschutz durch Technikgestaltung und datenschutzfreundliche Voreinstellungen, wie explizit durch Art. 25 DSGVO i. V. m. mit Erwägungsgrund 78 (Geeignete technische und organisatorische Maßnahmen) gefordert, ist zuvorderst ein kontinuierlicher Prozess. Werden demnach neue Technologien entwickelt, die nachweislich zeigen, dass Prinzipien des Datenschutzes mit angemessenem Aufwand besser als zuvor erreicht werden können, werden diese nach und nach auch zum neuen gesetzlich vorgeschriebenen Standard. Die hier vorgestellten technischen Beiträge demonstrieren, dass dies auch für die bislang eher im Hintergrund stehenden Prinzipien Transparenz und Auskunft (vgl. auch Pallas 2021) möglich ist.

# Danksagung

# Förderhinweis

# Literatur